\documentclass[conference]{IEEEtran}
\usepackage{amsmath,amssymb,graphicx}
\newtheorem{thm}{Theorem}[section]

\title{Cyclic Prefix Adaptation with Constant Overall Symbol Time for DFT-spread-OFDM and OFDM}

\author{\IEEEauthorblockN{Liangping Ma}
\IEEEauthorblockA{InterDigital Communications, Inc.\\
San Diego, CA 92121\\
Email: liangping.ma@interdigital.com}
\and
\IEEEauthorblockN{Tao Deng}
\IEEEauthorblockA{InterDigital Communications, Inc.\\
Melville, NY 11747\\
Email: tao.deng@interdigital.com}
\and
\IEEEauthorblockN{Alpaslan Demir}
\IEEEauthorblockA{InterDigital Communications, Inc.\\
Melville, NY 11747\\
Email: alpaslan.demir@interdigital.com}}

\begin{document}
\maketitle

\begin{abstract}
For DFT-spread-OFDM or OFDM, if the delay spread varies in a wide range and the symbol duration is relatively short, adapting the cyclic prefix (CP) duration rather than using a fixed one may significantly improve the spectral efficiency while preventing inter-symbol interference (ISI). In practice, it may be beneficial to have a constant overall DFT-spread-OFDM/OFDM symbol time, which is the sum of the duration of a CP and the duration of a data portion. We propose to adapt the CP duration to the delay spread without changing the overall symbol time for DFT-spread-OFDM or OFDM, and address implementation challenges. In particular, we propose changing the clocking rate of ADC and DAC or using a Farrow filter to reduce the computational complexity of arbitrary-size DFT/IDFT resulting from the adaptation.
\end{abstract}
%

%
\IEEEpeerreviewmaketitle

\section{Introduction}
\label{sec:intro}
Cyclic prefix (CP) has been widely used in practice to mitigate inter-symbol interference (ISI), for example, for OFDM in the UMTS long-term evolution (LTE) downlink and in IEEE 802.11a/g/n/ac, and for DFT-spread-OFDM in the LTE uplink. To fully eliminate ISI, the length of the CP should be at least as long as the delay spread. On the other hand, to maintain good spectral efficiency, the CP length cannot be too long.

It has long been noted that the delay spread may vary significantly from user to user, and from cell to cell. That motivated the definition of two CP lengths in LTE: a normal CP of length 4.7 $\mu$s and an extended CP of length 16.7 $\mu$s~\cite{Dhalman08}. The extended CP is intended to be used in environments of extensive delay spread (e.g., in large cells) or in the Multicast/Broadcast over Single Frequency Network (MBSFN) where the effective delay spread could be large due to the difference in propagation delays from different base stations.

Recent measurement studies show that for millimeter-wave wireless channels at 28GHz and 73GHz, the maximum root-mean-quare (RMS) delay spread (defined as the RMS of the power delay profile) could be tens of times greater than the average RMS delay spread and hundreds of times greater than the minimum~\cite{Sun14}. For the millimeter wave, the symbol duration tends to be short. Therefore, if we use a fixed CP length to prevent ISI, we have to make the CP long enough to accommodate the maximum RMS delay spread, resulting in very inefficient use of the transmission time. As an example, take the maximum RMS delay spread to be 200.3ns and the average 12.1ns at 73GHz from Table 2 of \cite{Sun14}. Let the subcarrier spacing be 0.5MHz, which implies a data portion of $2\mu s$ within an overall symbol time. We set the CP length to be a multiple, say, six times, of the RMS delay spread. This setup would ensure that 99.7\% of the delay lines to be covered by the CP if the delay lines follows a Gaussian distribution. To use a common CP length targeting the maximum, the CP would be 6 $\times$ 200.3ns and the overhead would be $1.2/(2+1.2)=37.5\%$. In contrast, the overhead could be reduced to 3.5\% if the CP length is set to 6 times of the actual RMS delay spread.



There has been a large amount of research on optimizing the CP duration for maximizing the spectral efficiency~\cite{Fet06}\cite{Tonello10}. However, to our knowledge, in the existing works the duration of a DFT-spread-OFDM/OFDM symbol, which in this paper includes a CP and a data portion as done in \cite[p. 322]{Dhalman08}, changes with the CP duration because the duration of the data portion is fixed, making it hard to compose fixed-duration frame structures while maintaining desired spectral efficiency. One needs to be aware that in the literature the DFT-spread-OFDM/OFDM symbol may refer to the data portion only.

In practice, it is desirable to have a fixed subframe duration for synchronous communication systems such as LTE. For one thing, it simplifies resource allocation and inter-cell interference management~\cite{Bera13} by having a common time unit. For another, it helps with backward compatibility with older systems that allocate network resource to a user in some basic time units, for example, the transmission time interval (TTI) of 2ms in HSPA and of 1 ms in LTE~\cite{Dhalman08}.

\begin{figure}[htb]
\begin{minipage}[b]{1.0\linewidth}
  \centering
  \centerline{\includegraphics[width=5cm]{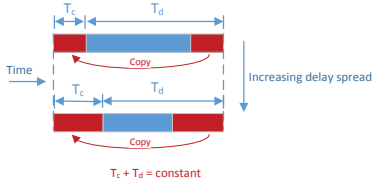}}
\end{minipage}
\caption{Adapt the CP duration while keeping the symbol duration constant.}
\label{fig:adapt}
\end{figure}

We propose \emph{Adaptive CP} to adapt the CP duration to the delay spread without changing the overall DFT-spread-OFDM or OFDM symbol time, as illustrated in Fig.~\ref{fig:adapt}, where $T_c$ is the CP duration and $T_d$ is the data portion duration. With Adaptive CP, a constant subframe duration is easily achieved and essentially there is no constraint on the granularity of the CP duration. It was considered infeasible in \cite{Bera13} to adapt the CP duration with fine granularity to the delay spread while meeting the constraint of a fixed subframe duration under the implicit assumption that $T_d$ is fixed. To see it, suppose that the subframe duration is equal to 500 $\mu$s and $T_d=$ 66.7$\mu$s as in LTE. Let $n$ be the integer number of symbols that fill up a subframe. Then $T_c$ must satisfy $n(66.7+T_c)=500$, which gives only two solutions 4.7 $\mu$s and 16.7 $\mu$s that have relatively low overhead among all possible solutions. In our present work, we remove the constraint that $T_d$ is fixed, and as a result we are able to achieve fine granularity in adapting the CP duration to the delay spread. However, removing the constraint also leads to challenges in system design and implementation, which we will address in this paper.

Besides the CP, the zero tail is also proposed for delay spread adaptation as in zero-tail DFT-spread-OFDM~\cite{Bera13}. However, the tails are not exactly zeros and are data dependent, and as a result exact cyclic convolution is not achieved, which may lead to significant bit error rate (BER) performance degradation at high SNR.

Unique Word (UW) is an alternative to mitigate ISI for OFDM~\cite{Huemer10} and in principle can be made to adapt to the delay spread. A redundant signal is added in the frequency-domain signal to generate a zero tail in the time-domain signal, and then a unique word is superposed on the zero tail. At the receiver, the channel-transformed unique word needs to be subtracted, and that requires an accurate estimate of the channel, which may be hard to do in practice. Recently, the idea of unique word is extended to DFT-spread-OFDM~\cite{Sahin15}.


The remainder of the paper is organized as follows. Section \ref{sec:sys} presents the adaptation scheme without regard to computational complexity, Section \ref{sec:implementation} proposes solutions to reduce the computational complexity. Lastly, Section \ref{sec:con} concludes the paper.

\section{Adapting the CP duration}
\label{sec:sys}
\subsection{Single user support}
\begin{figure}[htb]
\begin{minipage}[b]{1.0\linewidth}
  \centering
  \centerline{\includegraphics[width=8.5cm]{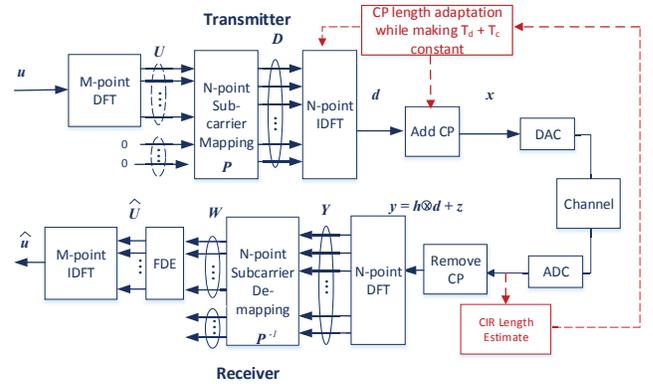}}  
\end{minipage}
\caption{System architecture for direct IDFT/DFT computation}
\label{fig:sys_DFT}
\end{figure}

The system architecture is shown in Fig.~\ref{fig:sys_DFT}. QAM symbols are fed to the system in blocks of length $M$. Consider an arbitrary block $\mathbf{u}=(u_{0}, u_{1}, \cdots, u_{M-1})^T$, where $^T$ stands for transpose. Let the output of the $M$-point DFT module be $\mathbf{U}=\mathrm{DFT}(\mathbf{u})=(U_{0}, U_{1}, \cdots, U_{M-1})^T$, where $U_k =\sum_{n=0}^{M-1} u_n e^{-j2\pi nk/M}$, where $k=0,1,\cdots, M-1$. Let $\mathbf{P}$ be an $N\times N$ permutation matrix used in subcarrier mapping. Let $\mathbf{0}_{1 \times (N-M)}$ be a $1 \times (N-M)$ vector with all entries being 0. The subcarrier mapping results in an $N$-vector $\mathbf{D}= \mathbf{P} (\mathbf{U}^T, \mathbf{0}_{1 \times (N-M)})^T $, which is fed to the $N$-point DFT module, resulting in $\mathbf{d}=\mathrm{IDFT}(\mathbf{D})=(d_0, d_1, \cdots, d_{N-1})^T$, where $d_k = \sum_{n=0}^{N-1} D_n e^{j2\pi nk/N}$, where $k=0,1,\cdots, N-1$. Let the channel impulse response (CIR) be $K+1$ samples long. The addition of a CP of $K$ samples results in the signal $\mathbf{x}=(d_{N-K}, d_{N-K+1}, \cdots, d_{N-1}, d_0, d_1, \cdots, d_{N-1})^T$ of length $(N+K)$, which is then passed to the DAC module, carrier modulated, and transmitted across the continuous-time channel which induces a discrete-time CIR $\mathbf{h}$. The received signal $\mathbf{y}=\mathbf{h} \otimes \mathbf{d}+ \mathbf{z}$, where $\otimes$ stands for circular convolution and $\mathbf{z}$ for noise. $\mathbf{Y} = \mathrm{DFT} (\mathbf{y})$, and $\mathbf{W}$ is equal to elements 1 through $M$ of $P^{-1}\mathbf{Y}$, where $P^{-1}$ is the inverse permutation. The equalization output is $\hat{\mathbf{U}}$, and the $M$-point IDFT output is $\hat{\mathbf{u}}$.

Now we consider how to determine the CP duration $T_c$ (in seconds) and the data portion duration $T_d$ (in seconds). The overall symbol time $T$ is chosen such that it is long enough to have a reasonably high efficiency $T_d/T$ while not being too long in order to satisfy other requirements such as limiting carrier-frequency offset and having an almost constant channel during $T$. The procedure of adapting $T_c$ to the delay spread is as follows. A statistic about the delay spread, for example the RMS delay spread $\tau$, is measured at the receiver, and fed back to the transmitter. Then, $T_c$ could be set as a multiple of $\tau$. Note that the discrete-time signal $\mathbf{x}$ enters the DAC module one sample per $T_s$, where $T_s$ is the period of the clocking signal of the DAC. Therefore, $T_c = K T_s$, yielding
\begin{equation}
K = T_c/T_s,
\label{eq:K}
\end{equation}
where we ignore the ceiling operation to simplify the notation. To maintain the same subframe duration, we keep the overall symbol duration $T=T_c + T_d$ constant. Thus, $T - T_c = N T_s$, or
\begin{equation}
N = (T-T_c)/T_s.
\label{eq:N}
\end{equation}
To achieve orthogonality among subcarriers, the subcarrier spacing $\Delta f = 1/(T-T_c)$. The bandwidth $B$ (in Hertz) of the DAC converted signal is
\begin{equation}
B =  N \Delta f = N/(T-T_c)= 1/T_s.
\label{eq:B}
\end{equation}
It follows from (\ref{eq:B}) and (\ref{eq:N}) that in order to keep $B$ the same, $T_s$ must remain the same and $N$ must be proportional to $T-T_c$.

To enable Adaptive CP for OFDM, we simply remove the $M$-point DFT and IDFT and set $M=N$ in Fig.~\ref{fig:sys_DFT}.

\subsection{Multiuser support}\label{sec:limit}
\begin{figure}[htb]
\begin{minipage}[b]{1.0\linewidth}
  \centering
  \centerline{\includegraphics[width=9cm]{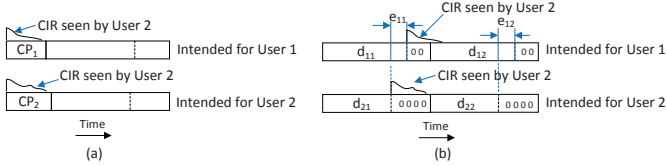}}
\end{minipage}
\caption{Transmission intended for user 1 interferes with user 2's receiving: (a) Adaptive CP, and (b) in zero-tail OFDM.}
\label{fig:int}
\end{figure}

In order to effectively support multiple users, the CP duration cannot be solely determined by the delay spread of individual users. Otherwise, multiple users on disjoint subcarriers with different CP durations may interfere with one another during simultaneous transmissions. This is shown in Fig.~\ref{fig:int}(a), where we consider the receiving of two simultaneous transmissions at user 2: one intended for user 1 and the other for user 2. Because the duration of CP$_1$ is shorter than that of CP$_2$, the superposed signal falling within user 2's DFT window is no longer circular, making the convolution non-circular.

 A similar problem exists with many other approaches such as \cite{Bera13}\cite{Huemer10}. In fact, as long as the CPs, zero-tails, or UWs are of different lengths, mutual interference may occur between users. As an example, Fig.~\ref{fig:int}(b) shows that for zero-tail OFDM~\cite{Huemer10} the fact that the two data segments $e_{11}$ and $e_{12}$ are almost always different breaks the cyclicity of the received signal in user 2's DFT window. Note that the CIRs shown in the figure are the ones seen by user 2. 

One approach to addressing the issue with Adaptive CP is to take the maximum of RMS delay spreads of only the users scheduled for simultaneous transmissions and configure these users with a common CP duration corresponding to the maximum. With this, significant gains could be achieved because the maximum of the delay spreads of a small number of users could be dramatically lower than that of all users in a cell.



A second approach is to configure a common CP duration for users of similar delay spreads and schedule only users of the same CP duration for simultaneous transmissions. A third approach is to use a filter to select only the desired frequencies for each intended receiver before doing DFT at the receiver. The idea of filtering on subcarriers or groups of subcarriers has been proposed for Filtered OFDM, Filter Bank Multicarrier~\cite{FBMC11} and resource block filtered OFDM~\cite{Li13}.

\section{reducing computational complexity}\label{sec:implementation}
We now present ways to reduce the complexity of the design in Section \ref{sec:sys}. 
The main complexity in Fig.~\ref{fig:sys_DFT} comes from DFT and IDFT, which have complexity about $N^2$ multiplications. The complexity is prohibitive for large $N$. However, if $N$ is a power of 2, we can use the efficient implementation radix-2 FFT, which has much lower complexity of about $(N/2) \log_2 N$ multiplications.
 On the other hand, if $N$ is not a power of 2, we need to explore other solutions. Mixed-radix IFFT/FFT (by factorizing $N$ into the powers of small prime numbers) has been considered in practice. However, the factorization changes with $N$, leading to changes in the hardware, which is undesirable in practice. We propose two solutions next.

%

\subsection{Changing the clocking rate for ADC and DAC}
\begin{figure}[htb]
\begin{center}
$ \begin{array}{c}
  \includegraphics[height=2.6cm]{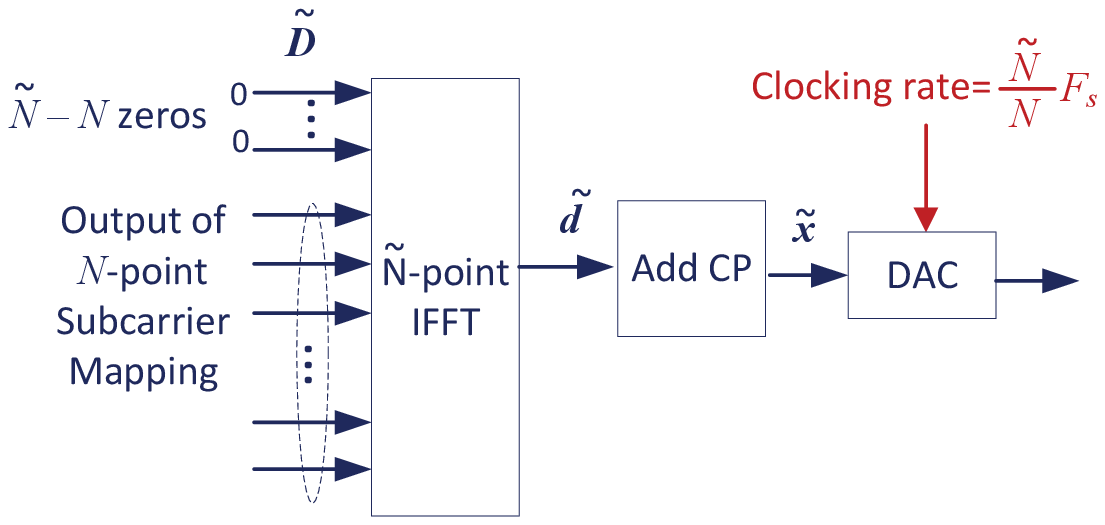} \\
  (a) \\
  \includegraphics[height=2.6cm]{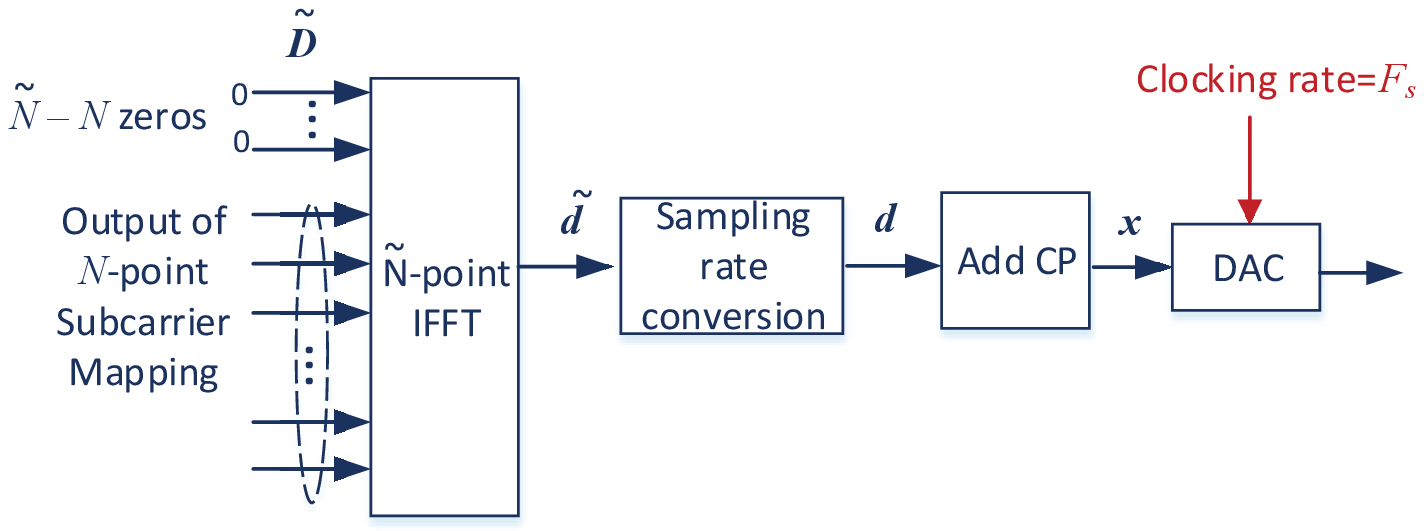} \\
  (b) \\
\end{array}
$
\end{center}
\caption{Transmitters with zero padding: (a) changing the clocking rate of the DAC and ADC; (b) sampling rate conversion.}
\label{fig:sys_FFT}
\end{figure}

We choose $\widetilde{N}$ to be a power of 2 and append zeros to $\mathbf{D}$ to get a length-$\widetilde{N}$ vector
\begin{equation}
\widetilde{\mathbf{D}}=(\mathbf{D}^T, \mathbf{0}_{1\times (\widetilde{N}-N)})^T
\label{eq:D_tilde}
 \end{equation}
and then apply $\widetilde{N}$-point radix-2 IFFT, as shown in Fig.~\ref{fig:sys_FFT}(a). Next, the IFFT output is clocked into the DAC at a rate $\widetilde{F}_s=\widetilde{N} F_s/N$ and denote the DAC output as $\widetilde{d}(t)$. Let the DAC output for the case of direct IDFT computation be $d(t)$. We claim that:
\begin{thm}
With the above zero-padding and clocking rate changing, $\widetilde{d}(t)=d(t)$.
\end{thm}
\noindent \emph{Proof:}
The IDFT output in Fig.~\ref{fig:sys_DFT} is
\begin{equation}
d(n)= \sum_{k=0}^{N-1} D_k e^{\frac{j2\pi kn}{N}} = \sum_{k=0}^{N-1} D_k e^{\frac{j2\pi kt}{N T_s}} \Big|_{t=nT_s}, \\0 \leq n \leq N-1.
\label{eq:d}
\end{equation}
Since $d(n)$ is fed into the DAC at rate $F_s=1/T_s$, by the Sampling Theorem the DAC output
\begin{equation}
d(t) = \sum_{k=0}^{N-1} D_k e^{j2\pi\frac{kt}{N T_s}}, \quad 0 \leq t \leq T- T_c.
\label{eq:dt}
\end{equation} This relationship is also explained in texts such as \cite{Stuber00}.

Now consider Fig.~\ref{fig:sys_FFT}(a). Define $\widetilde{T}_s=1/\widetilde{F}_s$. With (\ref{eq:D_tilde}), the $\widetilde{N}$-point IFFT output is
\begin{equation}
\widetilde{d}(n) = \sum_{k=0}^{N-1} D_k e^{j2\pi\frac{kn}{\widetilde{N}}}= \sum_{k=0}^{N-1} D_k e^{\frac{j2\pi kt}{\widetilde{N} \widetilde{T}_s}} \Big|_{t=n\widetilde{T}_s}, \quad 0 \leq n \leq \widetilde{N} - 1.
\end{equation}
Similarly, by the Sampling Theorem, we have
\begin{equation}
\widetilde{d}(t) = \sum_{k=0}^{N-1} D_k e^{j2\pi\frac{ kt}{\widetilde{N} \widetilde{T}_s}}, \\ \quad 0 \leq t \leq T- T_c.
\label{eq:dtt}
\end{equation}
Since $\widetilde{F}_s=\widetilde{N} F_s/N$, we have
\begin{equation}
\widetilde{N}\widetilde{T}_s = N T_s.
\label{eq:T_s_tilde}
\end{equation}
Comparing (\ref{eq:dt}) with (\ref{eq:dtt}), we have that $d(t)=\widetilde{d}(t)$. \hfill $\square$

Let the CP length in samples be $\widetilde{K}$, then $T_c/(T-T_c) = \widetilde{K}/\widetilde{N}$, which together with (\ref{eq:T_s_tilde}) yields
\begin{equation}
\widetilde{K} = T_c/\widetilde{T}_s.
\label{eq:K_tilde}
\end{equation}
The total bandwidth $\widetilde{B}$ is
\begin{equation}
\widetilde{B} = N/(T-T_c) = N/(\widetilde{N}\widetilde{T}_s).
\label{eq:B_tilde}
\end{equation}
\vskip -0.1cm


We now look at changing the clocking rate in practice. The $\widetilde{N}$-point IFFT output $\widetilde{d}(n)$ is fed to the DAC at a clocking rate $\widetilde{F}_s=\widetilde{N}/(N T_s)=\widetilde{N}/(T - T_c)$.
Thus the clocking rate $\widetilde{F}_s$ needs to change as $T_c$ adapts to the delay spread. Changing the clock rate is not new, and it has been used in frequency-hopping spread spectrum systems. In practice, the change in $\widetilde{F}_s$ can be achieved by the use of a programmable frequency synthesizer which typically uses one or more phase-locked loops (PLLs).

A performance metric particularly important to our problem is the PLL lock time, which is defined as the time that it takes the voltage-controlled oscillator (VCO) output to match the PLL reference clock signal in both frequency and phase, and it takes values on the order of tens of microseconds or less for modern wide-band frequency synthesizers~\cite{TI12}. Although the channel itself may change significantly on the order of milliseconds, the delay spread may change very little over a much longer time interval as the delay spread is primarily determined by the objects in the scattering environment within a beam. Under this assumption, the clocking rate of the DAC may not need to be changed frequently, therefore the loss of efficiency due to the change in clocking rate can be made negligible. In the case where the delay spread changes frequently, we may use two PLLs in a staggered fashion, one providing the clocking signal for the current communication, and the other to be turned on to provide a new clocking rate before the change in clocking rate is needed.

To account for the limited granularity of the clocking rate provided by a frequency synthesizer, we can work backwards from a set of $L$ available clocking rates $\widetilde{T}_s^{(1)}, \widetilde{T}_s^{(2)}, \dots, \widetilde{T}_s^{(L)}$ to determine $T_c$ by (\ref{eq:T_s_tilde}) and $NT_s=T-T_c$, $\widetilde{K}$ by (\ref{eq:K_tilde}), and $\widetilde{B}$ and $N$ by (\ref{eq:B_tilde}).



\subsection{Fractional sampling rate conversion}

\begin{figure}[htb]
\begin{minipage}[b]{1.0\linewidth}
  \centering
  \centerline{\includegraphics[width=2.5in]{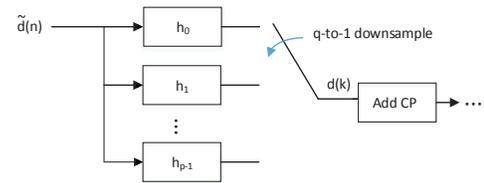}}
\end{minipage}
\caption{Polyphase decomposition for factor $p/q$ sampling rate conversion.}
\label{fig:poly}
\end{figure}

In this approach, we keep the clocking rate of the DAC (and ADC) at $F_s$, as shown in Fig.~\ref{fig:sys_FFT}(b). To ensure that $\widetilde{d}(t)$ is of duration $T- T_c$, we convert $\widetilde{d}(n)$, which corresponds to oversampling $d(t)$ at sampling frequency $\widetilde{F}_s=F_s\widetilde{N}/N$, into a shorter sequence $d(n)$ at a reduced sampling rate $F_s$. Polyphase filter decomposition~\cite{Proakis96} can be used, as shown in Fig.~\ref{fig:poly}, where $h_i$'s are the polyphase filters, $i=1,2, ..., p$. The complexity is reduced if $N/\widetilde{N}$ can be written as a ratio of two small integers $p$ and $q$ that are relatively prime, i.e., $N/\widetilde{N}=p/q$. For example, for $N=1536$ and $\widetilde{N}=2048$, we have $p=3$ and $q=4$. However, such simplification is not always available, especially if we want to have fine granularity in the CP duration adaptation. Additionally, when $N$ changes, the hardware structure for polyphase decomposition will change as well, which increases hardware complexity. This is clear because the original lowpass filter to be decomposed, which has a passband $[-1/(2\max(p,q)), 1/(2\max(p,q))]$ in relative frequency, changes with $p$ and $q$, and the number of polyphase filters (which is equal to $p$) changes with $p$.

\begin{figure}[htb]
\begin{minipage}[b]{1.0\linewidth}
  \centering
  \centerline{\includegraphics[width=3.9in]{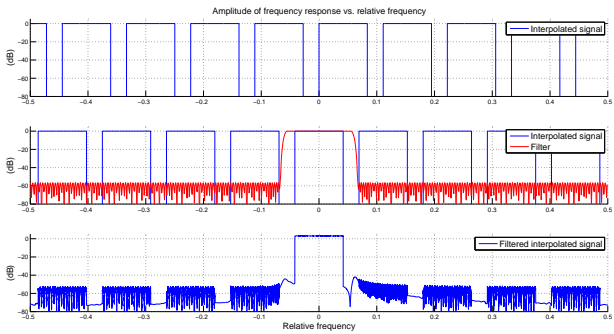}}
\end{minipage}
\caption{Amplitude of frequency response as a function of relative frequency for $N=1543$ and $\widetilde{N}=2048$ and an interpolation factor $p=9$. Top: interpolated signal; middle: shifted interpolated signal (blue line) and low pass filter (red line); bottom: filtered shifted interpolated signal.}
\label{fig:filter}
\end{figure}
\begin{figure}[htb]
\begin{minipage}[b]{1.0\linewidth}
  \centering
  \centerline{\includegraphics[width=3.7in]{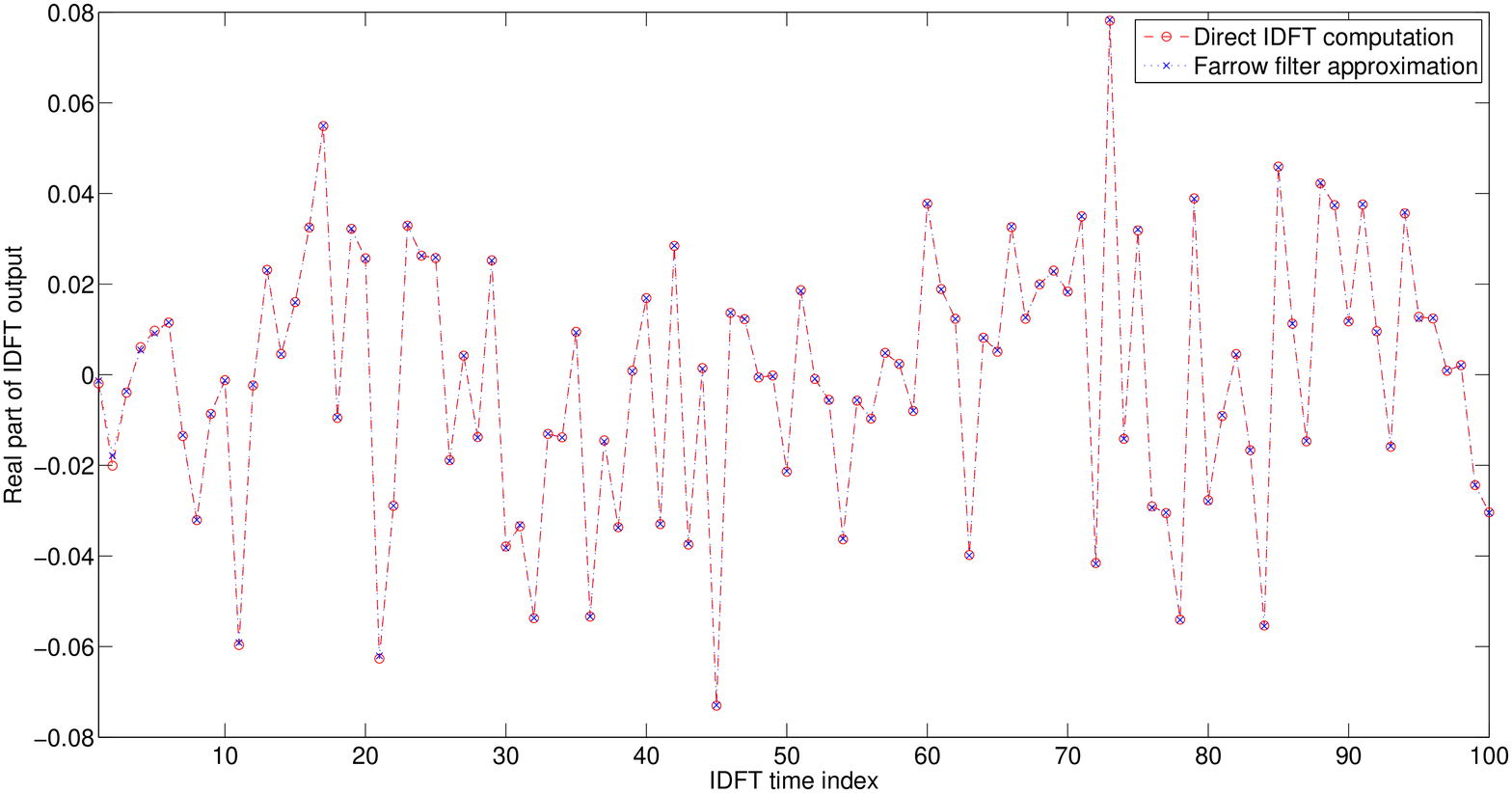}}
\end{minipage}
\caption{The real part of the IDFT output by direct computation (red circles with dashed line) and by Farrow approximation (blue crosses with dotted line) with order-4 polynomials.}
\label{fig:farrow}
\end{figure}

Alternatively, Farrow filter approximation~\cite[pp. 185-196]{Harris04} can be used to do arbitrary sampling rate conversion without changing the hardware structure. It works as follows. First, choose a constant integer $p$ and solve for $q=p\widetilde{N}/N$. Note that here $q$ may not be an integer any more. Then use lower order polynomials to approximate successive fragments of the original lowpass filter. Lastly `decimate' the output of the polyphase filters in strides of $q$, corresponding to a step size of $q/(p\widetilde{F}_s$) in seconds. A special treatment can significantly simplify the design. The original lowpass filter is symmetric in the frequency domain, but the spectrum of the output of the IFFT module is asymmetric with a support of $[0, N/\widetilde{N}]$ in relative frequency, resulting in a zero-interpolated signal with asymmetric spectrum with a support $[i/p, N/(p\widetilde{N})+i/p]$, $i=0, \pm 1, \dots $,  as illustrated in the top plot of Fig.~\ref{fig:filter}. To resolve this mismatch, we shift the spectrum of the IFFT output by multiplying $\widetilde{d}(n)$ with a phasor $\exp(-j\pi n N/\widetilde{N})$. This phase shift makes the spectrum of the interpolated signal symmetric in the frequency domain as shown in the middle plot of Fig.~\ref{fig:filter}. An inverse phase shift $\exp(j\pi n N/\widetilde{N})$ is applied to the output of the Farrow filter. The polynomial approximation can result in very good performance, as illustrated in Fig.~\ref{fig:farrow} for the first 100 data points of the IDFT/IFFT output. The relative MSE is -44.4dB, well below the effect of noise in a typical operating environment.

The Farrow approximation offers attractive complexity reduction. Assume that the original lowpass filter has a length $L$, and the polynomials are of order $\alpha$. Then, each polyphase filter has a length $\lceil L/p \rceil$. Using Horner's rule~\cite[p. 196]{Harris04}, the evaluation of a polynomial requires $\alpha$ multiplications. There are 2 multiplications for phase shifts for each sample. The total complexity is about $(\alpha + 1)\lceil L/p \rceil + 2 + \frac{\widetilde{N}}{2N} \log_2 \widetilde{N}$ multiplications per input sample, as opposed to $N$ in the direct IDFT/DFT approach. For the example in Fig.~\ref{fig:farrow}, $L=231, p=9, \alpha=4$. We have 146 multiplications per sample for the Farrow approximation method, as opposed to 1543 multiplications in the direct IDFT computation method.

\noindent \textbf{Note:} The Farrow approximation based approach described here offers a way to efficiently calculate arbitrary-size DFT or IDFT, and it can find many applications in practice.




\section{Conclusion}\label{sec:con}
We propose to adapt the CP duration to the delay spread without changing the overall symbol duration for DFT-spread-OFDM and OFDM to improve the spectral efficiency, and address the challenges in practical implementations. In particular, we propose changing the ADC/DAC clocking rate or computing arbitrary-size IDFT/DFT using Farrow approximation.

%

\noindent \textbf{Acknowledgement} The authors would like to thank Dr. Philip Pietraski, William Hackett, Dr. Robert A. DiFazio, Dr. Erdem Bala, Mihaela Beluri, Vincent Roy, Dr. Kyle Pan, Ravikumar Pragada and Dr. Shahrokh Nayeb Nazar of InterDigital for insightful discussions. The first author would also like to thank Prof. Fredric J. Harris of San Diego State University for providing the solution manual of his book~\cite{Harris04}.



%

\bibliographystyle{IEEEbib}

\end{document}